\documentclass[conference]{IEEEtran}
\IEEEoverridecommandlockouts
\usepackage{cite}
\usepackage{dblfloatfix}
\usepackage{overpic}
\usepackage{amsmath,amssymb,amsfonts}
\usepackage{graphicx}
\usepackage{textcomp}
\usepackage{xcolor}
\usepackage{float}
\usepackage{amsthm}
\usepackage{graphicx}
\usepackage{epstopdf}
\usepackage{amsmath,bm,stackengine}
\usepackage{amsfonts}
\usepackage{amssymb}
\usepackage{color}
\usepackage{multirow}
\usepackage{multicol}
\usepackage{soul,xcolor}
\usepackage{algorithm}
\usepackage{algpseudocode}

\theoremstyle{plain}

\newtheorem{lemma}{Lemma}

\newcommand{\vect}[1]{\mathbf{#1}}

\def\diag{\mathrm{diag}}

\def\tr{\mathrm{tr}}

\def\Htran{\mbox{\tiny $\mathrm{H}$}}
\def\Ttran{\mbox{\tiny $\mathrm{T}$}}
\def\CN{\mathcal{N}_{\mathbb{C}}} 
\def\imagunit{\mathsf{j}}

\begin{document}

\title{RIS-Assisted ISAC: Precoding and Phase-Shift Optimization for Mono-Static Target Detection 
\thanks{The work by \"O. T. Demir was supported by 2232-B International Fellowship for Early Stage Researchers Programme funded by the Scientific and Technological Research Council of T\"urkiye. The work by E. Bj{\"o}rnson was supported by the SUCCESS grant from the Swedish Foundation for Strategic Research.}}

\author{\IEEEauthorblockN{\"Ozlem Tu\u{g}fe Demir$^*$ and Emil Bj{\"o}rnson$^{\dagger}$}
\IEEEauthorblockA{{$^*$Department of Electrical-Electronics Engineering, TOBB University of Economics and Technology, Ankara, T\"urkiye
		} \\ {$^\dagger$Department of Computer Science, KTH Royal Institute of Technology, Stockholm, Sweden  
		} \\
		{Email:  ozlemtugfedemir@etu.edu.tr, emilbjo@kth.se\vspace{-0.3cm}}
}

}

\maketitle
\begin{abstract}
The reconfigurable intelligent surface (RIS) technology emerges as a highly useful component of the rapidly evolving integrated sensing and communications paradigm, primarily owing to its remarkable signal-to-noise ratio enhancement capabilities. In this paper, our focus is on mono-static target detection while considering the communication requirement of a user equipment. Both sensing and communication benefit from the presence of an RIS, which makes the channels richer and stronger. Diverging from prior research, we comprehensively examine three target echo paths: the direct (static) channel path, the path via the RIS, and a combination of these, each characterized by distinct radar cross sections (RCSs). We take both the line-of-sight (LOS) and the non-line-of-sight (NLOS) paths into account under a clutter for which the distribution is not known, but the low-rank subspace it resides.  We derive the generalized likelihood ratio test (GLRT) detector and introduce a novel approach for jointly optimizing the configuration of RIS phase-shifts and precoding. Our simulation results underscore the paramount importance of this combined design in terms of enhancing detection probability. Moreover, it becomes evident that the derived clutter-aware target detection significantly enhances detection performance, especially when the clutter is strong.
\end{abstract}

\section{Introduction}

While communication and radar technologies have traditionally advanced along separate trajectories, the recent paradigm shift known as ``integrated sensing and communications (ISAC)'' has emerged. ISAC aims to enhance cost-efficiency and power saving by sharing hardware resources and waveforms across communication, localization, and sensing applications \cite{Liu2022a}. In the pursuit of more energy- and hardware-efficient solutions, reconfigurable intelligent surfaces (RISs) have also gained prominence as a disruptive technology. RISs enable the control of wireless paths and the enhancement of the signal-to-noise ratio (SNR) and diversity. Incorporation of RIS technology into ISAC can not only strengthen wireless channel characteristics but also increase sensing performance, leading to improved SNR, accuracy, and reliability \cite{liu2023integrated,Elbir2023a}.

In the field of sensing, target detection plays a pivotal role, with the core objective to determine the presence or absence of a potential target in a designated direction or location \cite{buzzi2022foundations}. To enhance the performance of target detection, the utilization of multiple antennas at an ISAC transceiver with the support of RIS has gained significant attention \cite{Zhao2023, Luo2022}.

Although prior studies have delved into the concept of joint precoding at the transceiver and RIS phase-shift configuration, it is worth noting that some of them primarily centered on the role of RIS in aiding user equipments (UEs) for communication purposes \cite{Zhao2023, Luo2022}. On the contrary, \cite{zhang2022joint}  extended its scope to encompass both the target and communication UE channels, although with a focus on mutual information as opposed to an exhaustive analysis of target detection. In \cite{xing2023joint}, target detection is explored from the point of view of target size, but operated under the premise of the absence of a direct link between the potential target and the transceiver, resulting in the presence of a single echo originating from the cascaded channel assisted by RIS. Similarly, \cite{Liao2023} delved into the realm of a hybrid active-passive RIS-assisted ISAC system, assuming the absence of a static link within the sensing channel. In \cite{Luo2023}, the problem of joint precoding and RIS phase-shift configuration design is studied by considering all the static and RIS-aided cascaded paths. However, the work involves multiple UEs and targets, which makes the optimization problem very challenging and complicated. Instead, in \cite{meng2023ris}, simple algorithms are developed for a scenario with a single UE and single target. 

Distinguishing our work from the aforementioned studies and existing literature, we set out to investigate target detection via the utilization of three echo signals, each associated with its distinct radar cross-section (RCS) values. These echo paths correspond to different phenomena, including reflection through the line-of-sight (LOS) and non-line-of-sight (NLOS) paths connecting the ISAC transceiver and the target, the RIS-aided cascaded channel, and a hybrid mixture of both, occurring in cross directions. These coefficients become indispensable as the signal exhibits variability in its arrival and departure directions from the target.

Additionally, we assume the presence of clutter, modeled as a spatially correlated, target-free channel with unknown spatial statistics. Building on this comprehensive model of the sensing channel, we derive a generalized likelihood ratio test (GLRT) detector for binary hypothesis testing in target detection. The proposed detector leverages not only the statistics of the echo signals received through the LOS and NLOS  paths but also the low-rank subspace in which the clutter is known to reside. This approach enables clutter-aware superior performance compared to its clutter-unaware counterpart.

In addition to studying the target detection in detail, one of the major contributions of this paper is to propose a new precoding and RIS phase-shift optimization problem and derive a semi-closed-form solution to it. The problem aims to maximize the average channel gain of the novel sensing channel with three echo signals by satisfying the communication SNR requirement. The simulation results show significant gains in target detection when optimizing the RIS phase-shift configuration and ISAC precoding jointly.

\section{System and Channel Modeling}

\begin{figure}[t!]
        \centering 
	\begin{overpic}[width=0.8\columnwidth,tics=10]{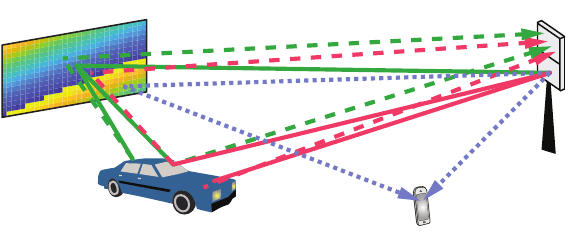}
\put(1,35){\small RIS}
\put(90,9){\small ISAC}
\put(84,5){\small transceiver}
\put(18,2){\small Target}
\put(65,2){\small UE}
 \end{overpic} 
        \caption{An ISAC transceiver with $K$ transmit antennas and $K$ receive antennas that want to transmit to a UE while also detecting the presence of a target. Both operations occur simultaneously with assistance from an RIS with $N$ elements. There are many paths in the channel. 
        Solid lines represent paths leading to the target and dashed lines are paths leading back to the receiver. Dotted lines represent paths to the UE.}
        \label{figure_mono_static_RIS}
        \vspace{-4mm}
\end{figure}
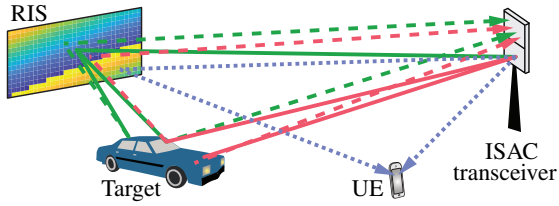

We consider an ISAC system that is deployed to guarantee a certain SNR at a single-antenna UE while detecting whether a target exists at a specific location. We consider the mono-static target detection scenario in Fig.~\ref{figure_mono_static_RIS} with downlink transmission to the UE.
The mono-static ISAC transceiver has $K$ transmit antennas and $K$ receive antennas, which are symmetrically arranged to achieve identical array response vectors.  An RIS with $N$ elements is deployed in the same area. The UE is served by the $K$-antenna transmitter with assistance from the RIS. 

In this paper, we assume the communication signals are also utilized for sensing without dedicated sensing symbols. However, the precoding design will take both sensing and communication requirements into account. Since data signals are used for sensing, we consider the transmission within a single time-frequency coherence block of a wideband system. Hence, the transmitted signals reach the receiver array within the same symbol time as transmitted.

\subsection{Communication channel modeling and performance}

We let $\vect{h}_{\textrm{s},1} \in  \mathbb{C}^K$ denote the static channel between the transmitter and the UE, that includes both LOS and NLOS paths. Furthermore, the cascaded channel via the RIS is represented by the vector
\begin{equation} \label{eq:cascaded-sensing-channel}
\vect{h}_{\textrm{c},1} =  \vect{a}_{\textrm{t}}  \vect{b}_{\textrm{t}}^{\Ttran}  \vect{D}_{\psi} \vect{h}_{\textrm{r},1} ,
\end{equation}
where $\vect{a}_{\textrm{t}}  \vect{b}_{\textrm{t}}^{\Ttran} \in \mathbb{C}^{K \times N}$ is the rank-one LOS channel matrix between the ISAC transceiver and RIS, $\vect{D}_{\psi}= \diag \left( e^{\imagunit \psi_1},\ldots, e^{\imagunit \psi_N} \right) \in \mathbb{C}^{N \times N}$ is the reflection matrix with continuous phase-shifts, and $\vect{h}_{\textrm{r},1} \in \mathbb{C}^N$ is the LOS+NLOS channel between the RIS and UE. The channel vectors $\vect{a}_{\textrm{t}}\in \mathbb{C}^K$ and $\vect{b}_{\textrm{t}}\in \mathbb{C}^N$ are the scaled array response vectors by the square roots of the channel gains corresponding to the LOS propagation between the transmitter and RIS.
All the UE channel coefficients are assumed to be known, so we can focus on the ISAC operation.\footnote{There are many channel estimation algorithms that work with sufficiently high accuracy \cite{Wei2021}.}
The end-to-end channel from the transmitter to the UE becomes
\begin{align}
    \vect{h}_1=  \vect{h}_{\textrm{s},1}+\vect{a}_{\textrm{t}}  \vect{b}_{\textrm{t}}^{\Ttran}  \vect{D}_{\psi} \vect{h}_{\textrm{r},1}.
\end{align}

We consider the transmission of an $L$-length sequence
$s[l]$, for $l=1,\ldots,L$, which consists of independent zero-mean unit-variance communication symbols.
These signals are transmitted using a precoding vector $\vect{p} \in \mathbb{C}^K$ that contains the transmit power and will be optimized. The received signal at the UE is
\begin{align}
    y_1[l] = \vect{h}_1^{\Ttran}\vect{p}s[l]+n_1[l],\quad l=1,\ldots,L, \label{eq:y1-l}
\end{align}
where the independent receiver noise is distributed as $n_1[l]\sim\CN(0,\sigma^2)$.
Hence, the communication SNR is
\begin{align}
    \textrm{SNR}_{1} = \frac{\left|\vect{h}_1^{\Ttran}\vect{p}\right|^2}{\sigma^2}.
\end{align}
To guarantee a certain quality-of-service, we require that $\textrm{SNR}_{1}\geq\gamma_{\rm th}$ for some predefined SNR threshold $\gamma_{\rm th}>0$.

\subsection{Mono-static target reflection modeling with unknown NLOS reflections}

The transmitted signal reaches the target location in two different ways: through the direct path and via the reflection by the RIS.
If the target exists, it will reflect the signals, and these can reach the receiver either through the direct LOS path, which is known, and NLOS paths, which are unknown, or via the RIS reflection in the same way. This gives rise to a total of four propagation channels from the transmitter to the receiver, which are indicated in Fig.~\ref{figure_mono_static_RIS}. We let $\overline{\vect{h}}_{\textrm{s},2} \in  \mathbb{C}^K$ denote the static LOS path between the transmitter and target location. Furthermore, the cascaded LOS channel via the RIS is represented by the vector
\begin{equation} \label{eq:cascaded-sensing-channel}
\overline{\vect{h}}_{\textrm{c},2} =  \vect{a}_{\textrm{t}}  \vect{b}_{\textrm{t}}^{\Ttran}  \vect{D}_{\psi} \overline{\vect{h}}_{\textrm{r},2} ,
\end{equation}
where $\overline{\vect{h}}_{\textrm{r},2} \in \mathbb{C}^N$ is the LOS channel between the RIS and target. 

We now let $\widetilde{\vect{h}}_{\textrm{s},2}$ and $\widetilde{\vect{h}}_{\textrm{r},2}$ denote the unknown NLOS channels between the transmitter and target, and between the RIS and target, respectively. Although these channels are unknown, their spatial correlation matrices are known, which are represented by $\vect{R}_{\textrm{s},2}$ and $\vect{R}_{\textrm{r},2}$, respectively. The cascaded NLOS channel via the RIS is denoted by $\widetilde{\vect{h}}_{\textrm{c},2} =  \vect{a}_{\textrm{t}}  \vect{b}_{\textrm{t}}^{\Ttran}  \vect{D}_{\psi} \widetilde{\vect{h}}_{\textrm{r},2} $ with the spatial correlation matrix $\vect{R}_{\textrm{c},2}=\vect{a}_{\textrm{t}}  \vect{b}_{\textrm{t}}^{\Ttran}  \vect{D}_{\psi}\vect{R}_{\textrm{r},2}\vect{D}^*_{\psi}\vect{b}^*_{\textrm{t}}\vect{a}^{\Htran}_{\textrm{t}} $.

If the target exists, the effective end-to-end channel to the receiver is given in \eqref{eq:effective-channel-target-detection} at the top of the next page,
\begin{figure*}
\begin{align} \label{eq:effective-channel-target-detection}
\vect{h}_2[l] =  \Bigg(& \underbrace{a_1[l] \left(\overline{\vect{h}}_{\textrm{s},2} \overline{\vect{h}}^{\Ttran}_{\textrm{s},2}+\overline{\vect{h}}_{\textrm{s},2} \widetilde{\vect{h}}^{\Ttran}_{\textrm{s},2}+\widetilde{\vect{h}}_{\textrm{s},2}\overline{\vect{h}}^{\Ttran}_{\textrm{s},2}\right)}_{\textrm{Direct path}} + \underbrace{a_2[l]
 \left(\overline{\vect{h}}_{\textrm{c},2} \overline{\vect{h}}_{\textrm{c},2}^{\Ttran}+\overline{\vect{h}}_{\textrm{c},2} \widetilde{\vect{h}}_{\textrm{c},2}^{\Ttran}+\widetilde{\vect{h}}_{\textrm{c},2} \overline{\vect{h}}_{\textrm{c},2}^{\Ttran}\right)}_{\textrm{Via surface}}
\nonumber \\ &+ \underbrace{a_3[l]\left(\overline{\vect{h}}_{\textrm{s},2}  \overline{\vect{h}}_{\textrm{c},2}^{\Ttran}+\overline{\vect{h}}_{\textrm{s},2}  \widetilde{\vect{h}}_{\textrm{c},2}^{\Ttran}+\widetilde{\vect{h}}_{\textrm{s},2}  \overline{\vect{h}}_{\textrm{c},2}^{\Ttran}\right)
+ a_3[l] \left(\overline{\vect{h}}_{\textrm{c},2} \overline{\vect{h}}_{\textrm{s},2}^{\Ttran}+\overline{\vect{h}}_{\textrm{c},2} \widetilde{\vect{h}}_{\textrm{s},2}^{\Ttran}+\widetilde{\vect{h}}_{\textrm{c},2} \overline{\vect{h}}_{\textrm{s},2}^{\Ttran}\right)}_{\textrm{Mix of direct and surface paths}}
\Bigg) \vect{p}.
\end{align}
\hrulefill
\vspace{-6mm}
\end{figure*}
where we included the precoding vector $\vect{p}$. Moreover, we have neglected the reflections from purely NLOS paths, as those signals are expected to be very weak due to the two-way path loss experienced by double NLOS paths. We assume that the channel $\vect{h}_2[l]$ follows a Gaussian distribution, motivated by the addition of many random paths and the central limit theorem.

The RCS is represented by three random realizations with zero mean. The variances of  $a_1[l]$, $a_2[l]$, and $a_3[l]$ are given by $\beta_1$, $\beta_2$, and $\beta_3$, respectively. Multiple RCS coefficients are required because the signal can reach and leave the target in different directions. However, the coefficient $a_3[l]$ appears twice due to channel reciprocity, which implies that the RCS is the same when the signal propagates from the transmitter to the target and back via the RIS, and when the signal travels in the opposite direction.
The four terms in \eqref{eq:effective-channel-target-detection} represent the four LOS+NLOS  propagation paths from the ISAC transmitter to the ISAC receiver. The first term is direct reflection by the target, which would also happen in the absence of the RIS. The second term is the path that reaches the target via the RIS and then goes back to the receiver in the same way. The third term is the path that reaches the target via the RIS and is then reflected through the direct path, while the fourth term takes the opposite direction.

\subsection{Clutter modeling}
Clutter can be modeled to exist in a low-rank subspace \cite{7041173}. We let it be spanned by the columns of the semi-unitary matrix $\vect{U}\in \mathbb{C}^{K\times r}$, where $r<K$ is the dimension of the subspace. Then, the clutter can be expressed as $\vect{U}\vect{x}[l]$, where the distribution of $\vect{x}[l]$ is not known. The subspace matrix $\vect{U}$ can be obtained by secondary data, where the ISAC transmitter keeps silent without transmitting anything and listens to the environment.

\section{Mono-Static Target Detection}

In this section, we will detail the considered target detection problem.
We denote the received signal at the ISAC receiver at time $l$ by $\vect{y}_2[l] \in \mathbb{C}^K$. We can then formulate the binary hypothesis test for target detection as
\begin{align}
&\mathcal{H}_0 \quad: \quad \vect{y}_2[l]=\vect{U}\vect{x}[l]+\vect{n}_2[l], \quad l=1,\ldots,L,  \label{eq:binary_hypothesis-RIS-swerling1}\\  
&\mathcal{H}_1 \quad: \quad \vect{y}_2[l]= \vect{h}_2[l] s[l]+\vect{U}\vect{x}[l]+\vect{n}_2[l], \quad l=1,\ldots,L,  \label{eq:binary_hypothesis-RIS-swerling1b}
\end{align}
where $\vect{n}_2[l] \sim \CN ( \vect{0}, \sigma^2 \vect{I}_K )$ is the independent additive noise vector.
The hypotheses represent the absence and presence of the target, respectively. We assume that self-interference is canceled to a great extent, and the residual self-interference is injected into the noise.

We will analyze the hypothesis test for fixed and known values of the precoding vector $\vect{p}$ and the RIS reflection matrix $\vect{D}_{\psi}$, and then optimize them in Section~\ref{sec:ISAC_optimization}.

If the hypothesis $\mathcal{H}_1$ is true, the desired channel's covariance matrix is  given as in \eqref{eq:R} at the top of the next page.
\begin{figure*}
\begin{align} \nonumber
\vect{R} &= \mathbb{E} \left\{ \vect{h}_2[l] \vect{h}_2^{\Htran}[l] \right\} 
  =
\beta_1   \left| \overline{\vect{h}}^{\Ttran}_{\textrm{s},2} \vect{p} \right|^2\left(  \overline{\vect{h}}_{\textrm{s},2}  \overline{\vect{h}}_{\textrm{s},2}^{\Htran}+\vect{R}_{\textrm{s},2}\right)+\beta_1\overline{\vect{h}}_{\textrm{s},2}  \overline{\vect{h}}_{\textrm{s},2}^{\Htran}\vect{p}^{\Htran}\vect{R}_{\textrm{s},2}^{\Ttran}\vect{p}+\beta_1\overline{\vect{h}}_{\textrm{s},2}  \overline{\vect{h}}_{\textrm{s},2}^{\Htran}\vect{p}^*\vect{p}^{\Ttran}\vect{R}_{\textrm{s},2}+\beta_1\vect{R}_{\textrm{s},2}\vect{p}^*\vect{p}^{\Ttran}\overline{\vect{h}}_{\textrm{s},2}  \overline{\vect{h}}_{\textrm{s},2}^{\Htran} \nonumber\\
&\quad+
 \beta_2   \left| \overline{\vect{h}}^{\Ttran}_{\textrm{c},2} \vect{p} \right|^2\left(  \overline{\vect{h}}_{\textrm{c},2}  \overline{\vect{h}}_{\textrm{c},2}^{\Htran}+\vect{R}_{\textrm{c},2}\right)+\beta_2\overline{\vect{h}}_{\textrm{c},2}  \overline{\vect{h}}_{\textrm{c},2}^{\Htran}\vect{p}^{\Htran}\vect{R}_{\textrm{c},2}^{\Ttran}\vect{p}+\beta_2\overline{\vect{h}}_{\textrm{c},2}  \overline{\vect{h}}_{\textrm{c},2}^{\Htran}\vect{p}^*\vect{p}^{\Ttran}\vect{R}_{\textrm{c},2}+\beta_2\vect{R}_{\textrm{c},2}\vect{p}^*\vect{p}^{\Ttran}\overline{\vect{h}}_{\textrm{c},2}  \overline{\vect{h}}_{\textrm{c},2}^{\Htran} \nonumber\\
&\quad+
 \beta_3 \left( \left(\overline{\vect{h}}_{\textrm{c},2}^{\Ttran} \vect{p}\right)  \overline{\vect{h}}_{\textrm{s},2} + \left(\overline{\vect{h}}_{\textrm{s},2}^{\Ttran} \vect{p}\right)  \overline{\vect{h}}_{\textrm{c},2} \right) \left( \left(\overline{\vect{h}}_{\textrm{c},2}^{\Ttran} \vect{p}\right)  \overline{\vect{h}}_{\textrm{s},2} + \left(\overline{\vect{h}}_{\textrm{s},2}^{\Ttran} \vect{p}\right)  \overline{\vect{h}}_{\textrm{c},2} \right)^{\Htran} \nonumber\\
 &\quad+\beta_3   \left| \overline{\vect{h}}^{\Ttran}_{\textrm{s},2} \vect{p} \right|^2\vect{R}_{\textrm{c},2}+\beta_3\overline{\vect{h}}_{\textrm{s},2}  \overline{\vect{h}}_{\textrm{s},2}^{\Htran}\vect{p}^{\Htran}\vect{R}_{\textrm{c},2}^{\Ttran}\vect{p}+\beta_3\overline{\vect{h}}_{\textrm{s},2}  \overline{\vect{h}}_{\textrm{s},2}^{\Htran}\vect{p}^*\vect{p}^{\Ttran}\vect{R}_{\textrm{c},2}+\beta_3\vect{R}_{\textrm{c},2}\vect{p}^*\vect{p}^{\Ttran}\overline{\vect{h}}_{\textrm{s},2}  \overline{\vect{h}}_{\textrm{s},2}^{\Htran} \nonumber\\
 &\quad+\beta_3   \left| \overline{\vect{h}}^{\Ttran}_{\textrm{c},2} \vect{p} \right|^2\vect{R}_{\textrm{s},2}+\beta_3\overline{\vect{h}}_{\textrm{c},2}  \overline{\vect{h}}_{\textrm{c},2}^{\Htran}\vect{p}^{\Htran}\vect{R}_{\textrm{s},2}^{\Ttran}\vect{p}+\beta_3\overline{\vect{h}}_{\textrm{c},2}  \overline{\vect{h}}_{\textrm{c},2}^{\Htran}\vect{p}^*\vect{p}^{\Ttran}\vect{R}_{\textrm{s},2}+\beta_3\vect{R}_{\textrm{s},2}\vect{p}^*\vect{p}^{\Ttran}\overline{\vect{h}}_{\textrm{c},2}  \overline{\vect{h}}_{\textrm{c},2}^{\Htran}. \label{eq:R}
\end{align}
\hrulefill
\vspace{-6mm}
\end{figure*}

Since the distributions of $\vect{x}[l]$ are unknown, we will use the GLRT detector.
In particular, $\vect{y}_2[l] -\vect{U}\vect{x}[l] \sim \CN(\vect{0}, \sigma^2\vect{I}_K)$ under the hypothesis $\mathcal{H}_0$ and
$\vect{y}_2[l]-\vect{U}\vect{x}[l]  \sim \CN(\vect{0},\vect{R}+\sigma^2\vect{I}_K)$ under the hypothesis $\mathcal{H}_1$.
 We should decide on the hypothesis $\mathcal{H}_1$ if the condition in \eqref{eq:likelihood-RIS}, shown at the top of the next page, holds,
 \begin{figure*}
\begin{align} \label{eq:likelihood-RIS}
\gamma &\leq \frac{\prod_{l=1}^L\underset{\vect{x}[l]}{\max}f_{\vect{y}_2[l]\big|\vect{x}[l],\mathcal{H}_1}\left(\vect{y}_2[l]\big|\vect{x}[l],\mathcal{H}_1\right)}{\prod_{l=1}^L\underset{\vect{x}[l]}{\max}f_{\vect{y}_2[l]\big|\vect{x}[l],\mathcal{H}_0}\left(\vect{y}_2[l]\big|\vect{x}[l],\mathcal{H}_0\right)} = \frac{\frac{1}{\pi^{LK} \left(\det \left( \vect{R}+\sigma^2\vect{I}_K\right)\right)^L } e^{-\sum_{l=1}^L\underset{\vect{x}[l]}{\min}(\vect{y}_2[l]-\vect{U}\vect{x}[l])^{\Htran}\left(\vect{R}+\sigma^2\vect{I}_K\right)^{-1}(\vect{y}_2[l]-\vect{U}\vect{x}[l])  }}{\frac{1}{\pi^{LK}\left( \det \left(\sigma^2\vect{I}_K\right)\right)^L } e^{-\sum_{l=1}^L\underset{\vect{x}[l]}{\min}(\vect{y}_2[l]-\vect{U}\vect{x}[l])^{\Htran}\left(\sigma^2\vect{I}_K\right)^{-1}(\vect{y}_2[l]-\vect{U}\vect{x}[l])}}.
\end{align}
\hrulefill
\vspace{-6mm}
\end{figure*}
 where $\gamma$ is the detector threshold that characterizes the detection probability and false alarm probability \cite{kay2009fundamentals}.  Here $f_{\vect{y}_2[l]|\vect{x}[l],\mathcal{H}_1}(.|.)$ and $f_{\vect{y}_2[l]|\vect{x}[l],\mathcal{H}_0}(.|.)$ denote the conditional probability density functions. The optimization problems in the numerator and denominator of \eqref{eq:likelihood-RIS} with respect to $\vect{x}[l]$ are quadratic minimization problems. After inserting the optimal values into it, we obtain the test detector as in \eqref{eq:likelihood-RIS2} at the top of the next page. Here, we introduced  the constant $c= \left(\det \left(\sigma^2\vect{I}_K\right)/\det \left(\vect{R}+\sigma^2\vect{I}_K\right)\right)^L$, which is independent of the received signals $\vect{y}_2[l]$.
 \begin{figure*}
 \begin{align}
 \label{eq:likelihood-RIS2}
\gamma  \leq c\frac{ e^{-\sum_{l=1}^L\left(\vect{y}_2^{\Htran}[l]\left(\vect{R}+\sigma^2\vect{I}_K\right)^{-1}\vect{y}_2[l]-\vect{y}_2^{\Htran}[l]\left(\vect{R}+\sigma^2\vect{I}_K\right)^{-1}\vect{U}\left(\vect{U}^{\Htran}\left(\vect{R}+\sigma^2\vect{I}_K\right)^{-1}\vect{U}\right)^{-1}\vect{U}^{\Htran}\left(\vect{R}+\sigma^2\vect{I}_K\right)^{-1}\vect{y}_2[l]\right)}}{ e^{-\sum_{l=1}^L\left(\sigma^{-2}\vect{y}_2^{\Htran}[l]\vect{y}_2[l]-\sigma^{-2}\vect{y}_2^{\Htran}[l]\vect{U}\vect{U}^{\Htran}\vect{y}_2[l]\right)}}.
 \end{align}
 \hrulefill
 \vspace{-6mm}
 \end{figure*}
 
We can rewrite the condition in \eqref{eq:likelihood-RIS2} by using the fact that $\ln(\gamma)$ is a monotonically increasing function for $\gamma\geq 0$:
\begin{align}  \nonumber
& \sum_{l=1}^L   \vect{y}_2^{\Htran}[l]\vect{T}\vect{y}_2[l] \geq \underbrace{(\ln(\gamma)-\ln(c))}_{=\gamma^{\prime}},  \label{eq:likelihood-RISb}
\end{align}
where $\gamma^{\prime}$ is the revised threshold variable that must be selected false alarm probability is set to a desired value. This value can be found by Monte Carlo trials. In this case, the sufficient statistics for target detection is $\sum_{l=1}^L   \vect{y}_2^{\Htran}[l]\vect{T}\vect{y}_2[l] $, where the matrix $\vect{T}$ is given as
\begin{align}
   \vect{T} =& \left(\vect{R}+\sigma^2\vect{I}_K\right)^{-1}\vect{U}\left(\vect{U}^{\Htran}\left(\vect{R}+\sigma^2\vect{I}_K\right)^{-1}\vect{U}\right)^{-1} \vect{U}^{\Htran}\nonumber\\
&\times\left(\vect{R}+\sigma^2\vect{I}_K\right)^{-1}  +\sigma^{-2}(\vect{I}_K-\vect{U}\vect{U}^{\Htran})-\left(\vect{R}+\sigma^2\vect{I}_K\right)^{-1}
\end{align}
and is affected by the precoding and surface configuration through $\vect{R}$.

\section{ISAC Precoding and Phase-Shift Optimization} \label{sec:ISAC_optimization}

The joint ISAC precoding vector $\vect{p}$ and RIS phase-shift configuration $\vect{D}_{\psi}$ will be optimized to maximize the gain of the cascaded sensing channel through the target while satisfying the communication SNR constraint $\textrm{SNR}_1\geq\gamma_{\rm th}$. The sensing channel gain can be expressed from \eqref{eq:R} as
\begin{align}
    \mathbb{E} \left\{ \| \vect{h}_2[l]  \|^2 \right\} = \tr ( \vect{R} ) = \vect{p}^{\Htran}\vect{C}\vect{p}
\end{align}
where the matrix $\vect{C}$ is computed in \eqref{eq:sensing-channel-gain} at top of the next page.

\begin{figure*}
    \begin{align} 
 \label{eq:sensing-channel-gain}
 \vect{C} & =\beta_1\left(\left\Vert\overline{\vect{h}}_{\textrm{s},2}\right\Vert^2+\tr\left(\vect{R}_{\textrm{s},2}\right)\right)\overline{\vect{h}}_{\textrm{s},2}^*\overline{\vect{h}}_{\textrm{s},2}^{\Ttran}+\beta_1\left\Vert\overline{\vect{h}}_{\textrm{s},2}\right\Vert^2\vect{R}_{\textrm{s},2}^{\Ttran}+\beta_1\overline{\vect{h}}_{\textrm{s},2}^*\overline{\vect{h}}_{\textrm{s},2}^{\Ttran}\vect{R}_{\textrm{s},2}^{\Ttran}+\beta_1\vect{R}_{\textrm{s},2}^{\Ttran}\overline{\vect{h}}_{\textrm{s},2}^*\overline{\vect{h}}_{\textrm{s},2}^{\Ttran} \nonumber\\
 &\quad + \beta_2\left(\left\Vert\overline{\vect{h}}_{\textrm{c},2}\right\Vert^2+\tr\left(\vect{R}_{\textrm{c},2}\right)\right)\overline{\vect{h}}_{\textrm{c},2}^*\overline{\vect{h}}_{\textrm{c},2}^{\Ttran}+\beta_2\left\Vert\overline{\vect{h}}_{\textrm{c},2}\right\Vert^2\vect{R}_{\textrm{c},2}^{\Ttran}+\beta_2\overline{\vect{h}}_{\textrm{c},2}^*\overline{\vect{h}}_{\textrm{c},2}^{\Ttran}\vect{R}_{\textrm{c},2}^{\Ttran}+\beta_2\vect{R}_{\textrm{c},2}^{\Ttran}\overline{\vect{h}}_{\textrm{c},2}^*\overline{\vect{h}}_{\textrm{c},2}^{\Ttran} \nonumber\\
 &\quad + \beta_3\left(\left\Vert\overline{\vect{h}}_{\textrm{s},2}\right\Vert^2+\tr\left(\vect{R}_{\textrm{s},2}\right)\right)\overline{\vect{h}}_{\textrm{c},2}^*\overline{\vect{h}}_{\textrm{c},2}^{\Ttran}+\beta_3\left\Vert\overline{\vect{h}}_{\textrm{s},2}\right\Vert^2\vect{R}_{\textrm{c},2}^{\Ttran}+\beta_3\overline{\vect{h}}_{\textrm{s},2}^*\overline{\vect{h}}_{\textrm{s},2}^{\Ttran}\vect{R}_{\textrm{c},2}^{\Ttran}+\beta_3\vect{R}_{\textrm{c},2}^{\Ttran}\overline{\vect{h}}_{\textrm{s},2}^*\overline{\vect{h}}_{\textrm{s},2}^{\Ttran}+\beta_3\overline{\vect{h}}_{\textrm{s},2}^{\Htran}\overline{\vect{h}}_{\textrm{c},2}\overline{\vect{h}}_{\textrm{c},2}^*\overline{\vect{h}}_{\textrm{s},2}^{\Ttran}\nonumber\\
 &\quad + \beta_3\left(\left\Vert\overline{\vect{h}}_{\textrm{c},2}\right\Vert^2+\tr\left(\vect{R}_{\textrm{c},2}\right)\right)\overline{\vect{h}}_{\textrm{s},2}^*\overline{\vect{h}}_{\textrm{s},2}^{\Ttran}+\beta_3\left\Vert\overline{\vect{h}}_{\textrm{c},2}\right\Vert^2\vect{R}_{\textrm{s},2}^{\Ttran}+\beta_3\overline{\vect{h}}_{\textrm{c},2}^*\overline{\vect{h}}_{\textrm{c},2}^{\Ttran}\vect{R}_{\textrm{s},2}^{\Ttran}+\beta_3\vect{R}_{\textrm{s},2}^{\Ttran}\overline{\vect{h}}_{\textrm{c},2}^*\overline{\vect{h}}_{\textrm{c},2}^{\Ttran}+\beta_3\overline{\vect{h}}_{\textrm{c},2}^{\Htran}\overline{\vect{h}}_{\textrm{s},2}\overline{\vect{h}}_{\textrm{s},2}^*\overline{\vect{h}}_{\textrm{c},2}^{\Ttran}.
 \end{align}
 \hrulefill
\end{figure*}
Then, the optimization problem can be cast as
\begin{subequations}\label{eq:optimization}
\begin{align} 
\underset{\vect{p},e^{\imagunit\psi_1},\ldots,e^{\imagunit\psi_N}}{\textrm{maximize}} \quad &\vect{p}^{\Htran}\vect{C}\vect{p} 
  \label{eq:optimization-obj}\\ 
\textrm{subject to}\,\,\,\quad &\left|\vect{h}_1^{\Ttran}\vect{p}\right|^2\geq \gamma_{\rm th}\sigma^2, \\
& \left\|\vect{p}\right\|^2 = P_{\rm t},
\end{align} 
\end{subequations}
where $P_{\rm t}$ is the total transmit power of the ISAC transmitter and the variables are the precoding vector
$\vect{p}$ and the entries of the RIS phase-shift matrix $\vect{D}_{\psi}$.
The problem \eqref{eq:optimization} is hard to solve jointly and globally since it is non-convex. There are several factors that contribute to the non-convexity. First, the unit-modulus constraints on the RIS responses and maximizing a convex quadratic function make the problem non-convex. The communication SNR constraint is also non-convex. Last but not least, the precoding vector and the phase-shift variables are coupled in the objective function and the communication SNR constraints. Therefore, we will first select $e^{\imagunit\psi_1},\ldots,e^{\imagunit\psi_N}$ to maximize $\|\overline{\vect{h}}_{{\rm c},2}\|^2$ that appear multiple times in $\vect{C}$ in \eqref{eq:sensing-channel-gain} heuristically. Later, we will propose a novel semi-closed-form solution for the 
 precoding vector $\vect{p}$ that solves the above problem while keeping the phase-shifts constant.

To maximize $\|\overline{\vect{h}}_{{\rm c},2}\|^2$, the phase-shift should be selected to maximize $|\vect{b}_{\textrm{t}}^{\Ttran}  \vect{D}_{\psi} \overline{\vect{h}}_{\textrm{r},2}|^2$ from \eqref{eq:cascaded-sensing-channel}, which is achieved by 
\begin{equation} 
\psi_n = \angle\left(b_{\textrm{t},n}^*\overline{h}_{\textrm{r},2,n}^*\right) ,\label{eq:psi-n}
\end{equation}
where $b_{\textrm{t},n}$ and $\overline{h}_{\textrm{r},2,n}$ denote the $n$-th entries of the vectors $\vect{b}_{\textrm{t}}$ and $\overline{\vect{h}}_{\textrm{r},2}$, respectively, and $\angle(\cdot)$ gives the phase of its argument.

\subsection{Precoding optimization for given phase-shift configuration} \label{sec:precoding}

Denoting that the rank of $\vect{C}$ by $r_2$, we can decompose the positive semi-definite matrix $\vect{C}$ as $\vect{C}=\vect{B}\vect{B}^{\Htran}$ for some $\vect{B}\in \mathbb{C}^{K \times r_2}$. We will assume $K\geq r_2+1$ and construct the matrix
\begin{align} \label{eq:A}
    \vect{A} = \begin{bmatrix} \vect{h}_1^* & \vect{B}\end{bmatrix}.
\end{align}
Its QR decomposition can be expressed by using Gram-Schmidt orthogonalization as\footnote{Without loss of generality, we assume the rank of $\vect{A} $ is $r_2+1$. For the rank-deficient case, a simpler and similar methodology can be applied to find the optimal precoding vector.}
\begin{align} \label{eq:A-QR}
   \vect{A}= \vect{Q}_{\rm A} \vect{R}_{\rm A},
\end{align}
where the columns of $\vect{Q}_{\rm A}\in \mathbb{C}^{K \times (r_2+1)}$ construct an orthonormal set of vectors with its first column being $\vect{h}_1^*/\|\vect{h}_1\|$. The entries of the upper triangular matrix $\vect{R}_{\rm A}$ are denoted as
\begin{align}  \label{eq:A-R}
\vect{R}_{\rm A} = \begin{bmatrix} \|\vect{h}_1\| & \overline{\vect{r}}_{\rm A}^{\Htran}  \\
\vect{0} & \overline{\vect{R}}_{\rm A} 
 \end{bmatrix} ,
    \end{align}
    where $\overline{\vect{r}}_{\rm A}$ and $\overline{\vect{R}}_{\rm A}$ are obtained from   Gram-Schmidt orthogonalization. It can be shown that the optimal precoding vector should lie in the subspace spanned by the columns of $\vect{Q}_{\rm A}$, hence, it can be expressed as $\vect{p}=\vect{Q}_{\rm A}\vect{x}$ in terms of some $\vect{x}\in \mathbb{C}^{r_2+1}$.\footnote{If the optimal precoding vector had a component in the null-space of $\vect{Q}_{\rm A}$, and, thus $\vect{A}$, then that component would affect neither objective function nor the communication constraint, just resulting in a loss of power. Hence, the optimal precoding vector should lie in the subspace spanned by the target and UE channels.} We note that
    \begin{align}
        \vect{p}^{\Htran}\vect{A} &= \begin{bmatrix} \vect{p}^{\Htran}\vect{h}_1^* & \vect{p}^{\Htran}\vect{B}\end{bmatrix}  = \vect{x}^{\Htran}\vect{Q}_{\rm A}^{\Htran}\vect{Q}_{\rm A}\vect{R}_{\rm A} = \vect{x}^{\Htran}\vect{R}_{\rm A},
    \end{align}
    where we used $\vect{Q}_{\rm A}^{\Htran}\vect{Q}_{\rm A}=\vect{I}_{r_2+1}$. For a fixed RIS configuration, we can then recast the problem in \eqref{eq:optimization} in terms of $\vect{x}$ as
\begin{subequations}
\begin{align}
\underset{\vect{x}}{\textrm{maximize}} \quad &\vect{x}^{\Htran}\vect{F}\vect{F}^{\Htran}\vect{x}\\ 
\textrm{subject to}\quad &x_1\geq \frac{\sqrt{\gamma_{\rm th}\sigma^2}}{\|\vect{h}_1\|}, \label{eq:first-constraint} \\
& \|\vect{x}\|^2= P_{\rm t}, \\
& x_1\in \mathbb{R}, \quad x_1\geq 0,
\end{align}
\end{subequations}
 where $\vect{F}=\left[\overline{\vect{r}}_{\rm A} \ 
 \overline{\vect{R}}_{\rm A}^{\Htran}\right]^{\Htran}$ and we set $x_1$ as a non-negative real-valued variable without changing optimality of the original problem. From the Karush-Kuhn-Tucker (KKT) conditions, there are two cases to be evaluated for the candidate optimal solution $\vect{x}$:

{\bf Case a:} In the first case, the first constraint in \eqref{eq:first-constraint} is satisfied with a strict inequality and we look for $\vect{x}$ that maximizes $\vect{x}^{\Htran}\vect{F}\vect{F}^{\Htran}\vect{x}$ under the constraints that $\|\vect{x}\|^2=P_{\rm t}$ and $x_1\geq 0$. The solution to this problem is the phase-shifted and scaled eigenvector of $\vect{F}\vect{F}^{\Htran}$ corresponding to the largest eigenvalue so that $\|\vect{x}\|^2=P_{\rm t}$ and $x_1\geq 0$. If the resulting $x_1\geq \frac{\sqrt{\gamma_{\rm th}\sigma^2}}{\|\vect{h}_1\|} $, then the found $\vect{x}$ is the optimal solution. 

{\bf Case b:} In the second case, we check the boundary condition for the first constraint where it is satisfied with the equality $x_1= \frac{\sqrt{\gamma_{\rm th}\sigma^2}}{\|\vect{h}_1\|}$. Under this condition, the optimal $\overline{\vect{x}}=[x_2 \ \ldots \ x_{r_2+1}]^{\Ttran}$ can be found by solving the problem
\begin{subequations} \label{eq:optimization3}
\begin{align}
\underset{\overline{\vect{x}}}{\textrm{maximize}} \quad & \overline{\vect{x}}^{\Htran}\overline{\vect{A}}\overline{\vect{x}}+2\Re\left(\overline{\vect{x}}^{\Htran}\overline{\vect{b}}\right)\\ 
\textrm{subject to}\quad 
& \|\overline{\vect{x}}\|^2= \overline{c}, 
\end{align}
\end{subequations}
where 
\begin{align}
&\overline{\vect{A}}=
\overline{\vect{R}}_{\rm A}\overline{\vect{R}}_{\rm A}^{\Htran}, \\
& \overline{\vect{b}}=\frac{\sqrt{\gamma_{\rm th}\sigma^2}}{\|\vect{h}_1\|}  \overline{\vect{R}}_{\rm A}\overline{\vect{r}}_{\rm A}, \\
&\overline{c}= P_{\rm t}-\frac{\gamma_{\rm th}\sigma^2}{\|\vect{h}_1\|^2}.
\end{align}

The above problem can be solved using a similar method as in \cite[App.~A]{is_channel}, where the only difference is that the inequality constraint in \cite[(47b)]{is_channel} is replaced by equality. 

\begin{lemma}\label{lemma:norm_maximization}
	Let $\lambda_{1}\geq \cdots\geq \lambda_{r_2}\geq0$ be the nonnegative eigenvalues and $\vect{u}_{i}\in \mathbb{C}^{r_2}$, for $i=1,\ldots,r_2$  be the corresponding orthonormal eigenvectors of $\overline{\vect{A}}$.
If $\overline{\vect{b}}=\vect{0}$, the optimal solution to the problem \eqref{eq:optimization3} is given by $\overline{\vect{x}}^{\star}=\sqrt{\overline{c}}\vect{u}_{1}$. Otherwise, the optimal solution is  
\begin{equation}
\overline{\vect{x}}^{\star} = \sum_{i=1}^{r_2}\frac{\vect{u}_{i}\vect{u}_{i}^{\Htran}\overline{\vect{b}}}{\gamma^{\star}-\lambda_{i}}, \label{eq:phi}
\end{equation}
where $\gamma^{\star}>\lambda_1$ is the unique root of
\begin{equation}
\sum_{i=1}^{r_2}\frac{\left\vert\vect{u}_{i}^{\Htran}\overline{\vect{b}}\right\vert^2}{\left(\gamma-\lambda_{i}\right)^2}=\overline{c},\label{eq:root2}
\end{equation} which can be solved using a bisection search.
\end{lemma}

\section{Numerical Results}

In this section, we will demonstrate how the target sensing performance depends on the system parameters and precoding/phase-shift optimization.
The ISAC transceiver is equipped with a half-wavelength-spaced uniform planar array with $K=36$ transmit and receive antennas. We let $\varphi$ and $\theta$ denote the azimuth and elevation angles, respectively.
The communication UE is located in the direction $(\varphi,\theta)=(\pi/3,-\pi/5)$ seen from the transceiver while the target location is in the direction $(\varphi,\theta)=(\pi/6,-\pi/5)$. The RIS is deployed as a uniform square array with  $8\times 8$ half-wavelength-spaced elements. The UE is seen in the direction $(\varphi,\theta)=(0,-\pi/5)$, while the target is seen in the direction $(\varphi,\theta)=(-\pi/4,-\pi/5)$ from the RIS. The RIS is located in the direction $(\varphi,\theta)=(-\pi/4,0)$ seen from the transceiver, and the transceiver is located in the direction $(\varphi,\theta)=(\pi/4,0)$ seen from the RIS. The relative channel gain through a single RIS element in comparison to the static channel between the transceiver and UE or target is $-10$\,dB. The ratio of the UE channel gain multiplied by the transmit power to the noise variance is $0$\,dB and the SNR requirement of the UE is $\gamma_{\rm th}=10$; 
 thus, the communication performance is guaranteed throughout the simulations. The number of symbols used for target detection is $L=5$. The RCS variance for the reflection through the direct path, surface, and mix of direct and surface paths is $\beta_1=0.1$, $\beta_2=4$, and $\beta_3=1$, respectively.

 The spatial correlation matrices for the communications and sensing channels are generated using the local scattering model in  \cite{demir2022channel} with six clusters centered around the $40^{\circ}$ and $20^{\circ}$ neighborhoods of the nominal azimuth and elevation angles, respectively with $10^{\circ}$ angular standard deviation. The same local scattering model is used to characterize the correlation matrix of the clutter with six clusters centered around the $20^{\circ}$ and $10^{\circ}$ neighborhoods of the nominal azimuth and elevation angles, respectively, with $5^{\circ}$ angular standard deviation. The nominal angles of the clutter are the same as the angles of the RIS seen by the BS.

\begin{figure}[t!]
        \centering 
	\begin{overpic}[width=0.8\columnwidth,tics=10]{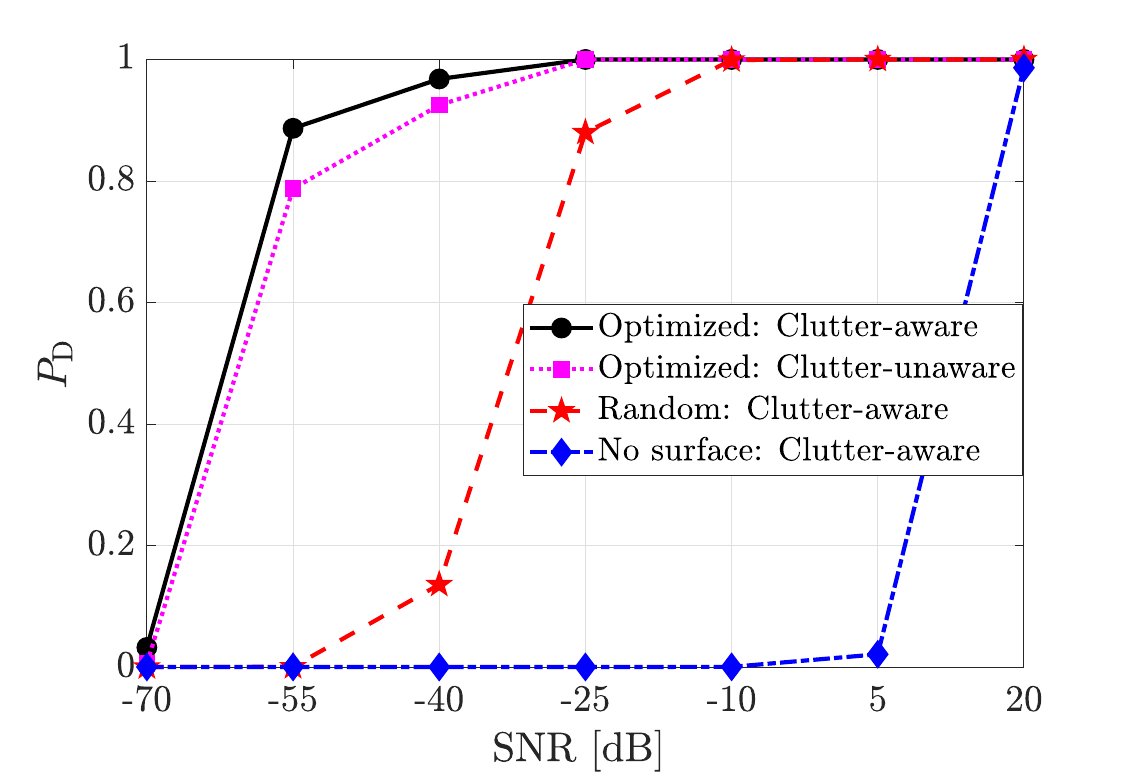}
 \end{overpic} 
  \vspace{-3mm}
        \caption{The detection probability with respect to the SNR for the received clutter-noise ratio of $20$\,dB.}
        \label{figure3} 
        \vspace{-4mm}
\end{figure}

Fig.~\ref{figure3} shows the detection probability, $P_{\rm D}$, versus the SNR of the two-way static target channel, where the SNR of the channel through RIS is $-10-10=-20$\,dB lower since the relative channel gain through RIS is $-10$\,dB lower as specified before. The received clutter's signal power is 20\,dB higher than the noise variance. The clutter-aware GLRT detector in \eqref{eq:likelihood-RISb} and the clutter-unware detector (which is obtained by setting $\vect{U}=\vect{0}$) are used with the false alarm probability $P_{\rm FA}=\alpha=10^{-3}$, from which the detector threshold is determined through multiple setups. The dash-dotted blue curve with diamonds represents the detection performance without the RIS, in which case the optimal precoding is found by Lemma~1. The dashed red curve with stars is obtained when the RIS is added to the setup, but it has a random configuration (i.e., it approximates a diffusely scattering surface) while only the precoding is optimized.
For both of these cases, the clutter-aware detector is utilized. The enhanced SNR provided by the RIS is evident in the improved detection probability. However, the improvement becomes significantly more pronounced when the precoding and RIS phase-shifts are jointly optimized using \eqref{eq:psi-n} and Lemma~1. These results are depicted by the solid black curve with circles (``Optimized''), which practically reaches a value of 1 for SNRs exceeding $-40$\,dB.

The ``optimized'' scheme with the clutter-unaware detector is illustrated by the dotted magenta curve with square markers. As expected, the detection performance is degraded in comparison to that of the clutter-aware detector, which leverages knowledge of the clutter subspace.

\begin{figure}[t!]
        \centering 
	\begin{overpic}[width=0.8\columnwidth,tics=10]{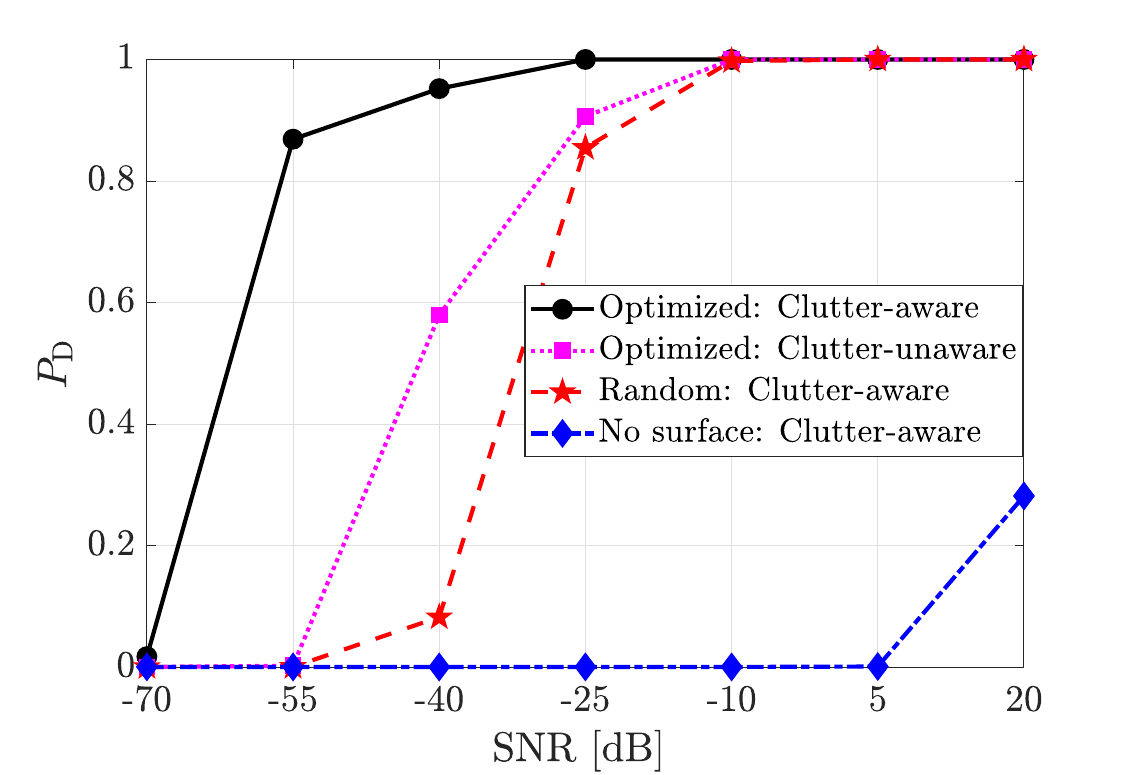}
 \end{overpic} 
 \vspace{-3mm}
        \caption{The detection probability with respect to the SNR for the received clutter-noise ratio of $40$\,dB.}
        \label{figure4} 
        \vspace{-4.5mm}
\end{figure}

In Fig.~\ref{figure4}, we replicate the earlier experiment, this time with an increase in the received clutter-noise ratio from $20$\,dB to $40$\,dB. This amplifies the clutter power, and all the clutter-aware schemes only slightly degrades, while the clutter-unaware one deteriorates significantly.

\section{Conclusions}

We have investigated a mono-static ISAC setup, where a single transceiver simultaneously serves a communication UE and detects the presence of a target at a specific location. Both the sensing and communication tasks are assisted by an RIS in the environment. Distinguishing our work from existing studies, we have considered all three potential target echo reflections with LOS and NLOS paths. We have derived the clutter-aware GLRT detector, assuming knowledge of the low-rank subspace of the clutter. 

We have additionally introduced a novel approach for jointly optimizing the configuration of RIS phase-shifts and precoding, based on the combined two-way sensing channel. Our simulation results clearly highlight the substantial improvements achieved through the proposed joint design, as compared to randomly adjusting the RIS phase-shift configurations or operating without an RIS. Additionally, the clutter-aware detector exhibits superior performance compared to the clutter-unaware detector, especially in scenarios where the clutter power is significantly strong.

\bibliographystyle{IEEEtran}
\bibliography{IEEEabrv,refs}

\end{document}